\documentstyle[12pt,psfig]{article}
\begin{document}

\hfill DUKE-CGTP-2001-04

\hfill hep-th/0103008

\vspace*{1.5in}

\begin{center}

{\large\bf Quotient Stacks and String Orbifolds }

\vspace{1in}

Eric Sharpe \\
Department of Physics \\
Box 90305 \\
Duke University \\
Durham, NC  27708 \\
{\tt ersharpe@cgtp.duke.edu} \\

$\,$ \\

\end{center}

In this short review we outline some recent developments in
understanding string orbifolds.  In particular, we outline the
recent observation that string orbifolds do not precisely describe
string propagation on quotient spaces, but rather are {\it literally}
sigma models on objects called quotient stacks, which are closely
related to (but not quite the same as) quotient spaces.  
We show how this is an immediate
consequence of definitions, and also how this explains a number of
features of string orbifolds, from the fact that the CFT is well-behaved
to orbifold Euler characteristics.  Put another way, many features
of string orbifolds previously considered ``stringy'' are now understood
as coming from the target-space geometry; one merely needs to identify
the correct target-space geometry. 

\begin{flushleft}
February 2001 
\end{flushleft}

\newpage

\tableofcontents

\newpage

\section{Introduction}

\begin{figure}
\centerline{\psfig{file=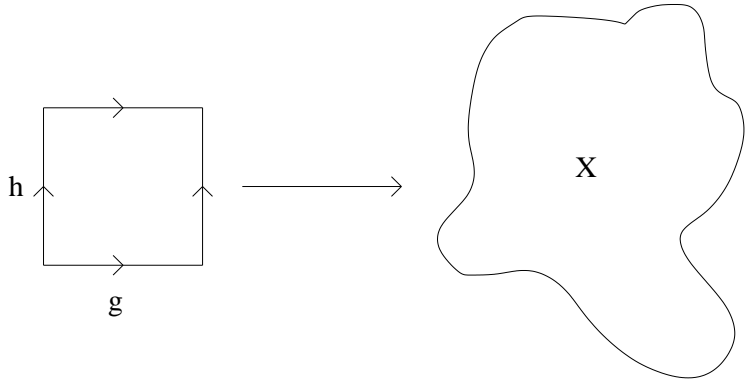,width=4in}}
\caption{\label{figsig1} Contribution to the $(g,h)$ twisted sector
of a string orbifold on $T^2$}
\end{figure}

One often hears that string orbifolds \cite{strorb1,strorb2,dixonthes}
describe strings propagating
on quotient spaces (with some `stringy' effects
at singularities).  Of course, a string orbifold is not 
described in terms of maps into a quotient space, but rather
is set up in terms of group actions on covering spaces.
For example, the partition function of a string orbifold is
of the form
\begin{displaymath}
Z \: \propto \: \sum \, Z\left( \mbox{twisted sector} \right)
\end{displaymath}
e.g., the partition function of a string orbifold on $T^2$ can be
written
\begin{displaymath}
Z_{T^2} \: = \: \frac{ 1 }{ |G| } \sum_{g, h, gh = hg} \, Z_{g,h}
\end{displaymath}
where each $Z_{g,h}$ is a sigma model from a square into $X$,
such that the sides of the square are related by the action of $g, h
\in \Gamma$, as illustrated in figure~\ref{figsig1}.

We shall argue later that, in fact, the twisted sector sum and the functional
integral within each twisted sector of 
a string orbifold partition function are are both duplicated
by a sum over maps into
something known as a quotient {\it stack}, an object which is
closely related to the quotient space.  Quotient stacks
look mostly like quotient spaces, but
have ``extra information'' over
any singularities of the quotient space, which makes
quotient stacks much better behaved than quotient spaces.
The fact that a sum over maps into a quotient stack duplicates the sums
in a string orbifold is a smoking gun for an interpretation of string
orbifolds as sigma models on quotient stacks.  We shall also outline
how one can define a classical action for a sigma model on a stack, which 
reproduces string orbifolds in the special case of quotient stacks.


Of course, this is a rather strong conclusion to most physicists.
Mathematicians, on the other hand, may find this result
considerably more natural.  After all, one way of thinking about
quotient stacks is as extraordinarily overcomplicated ways of describing
group actions on covers, which is the language used in string
orbifolds.  Also, quotient stacks have the structure of
a `generalized space' -- many stacks can be considered to be
slight generalizations of ordinary spaces, just as
noncommutative geometry gives a way of generalizing ordinary
geometry.  In particular, one often hears\footnote{Readable
references describing, in detail, differential geometry on stacks
are hard to find.  For this reason, reference \cite{qstx} included
a lengthy discussion of differential geometry on stacks.} that
one can do differential geometry on stacks, a necessary
(but by no means sufficient) condition to be able to compactify
a string.


Although we shall describe classical actions for strings on stacks,
generalizing string orbifold actions, and describe how a sum over
maps into a quotient stack duplicates both the twisted sector and
functional integral sums in a string orbifold, we should emphasize
that a tremendous amount of work remains to be done to check whether
the notion of string compactification on stacks is indeed sensible.
We have only set up the basics in \cite{qstx}.

One question one might well ask is simply, why bother?
If at the end of the day, we are merely using quotient stacks
as a radically overcomplicated way of describing group actions
on covers, then there is hardly a point.  Part of the reason
for our interest in string orbifolds specifically is that
this approach seems to give a new geometric understanding of
certain physical features of string orbifolds.  Another reason
is that this has nontrivial implications for the role string
orbifolds play in the rest of string theory -- for example,
constructions of moduli spaces of string vacua may need to
be slightly rethought, considering that deformation theory
of a string orbifold would have to be understood in terms of
deformation theory of a quotient stack, instead of a quotient
space.


The work outlined herein was originally done in an attempt to
understand whether it is possible to compactify a string on
a stack (here thought of as a slight generalization of a space).
One sometimes hears that, for example, one can perform
differential geometry on stacks, which would be one necessary
condition; however, a tremendous amount of work must be done
to check whether this notion is sensible.  We shall outline
the beginnings of a program to understand string compactification
on stacks -- specifically, we shall describe sigma models on stacks.
Now, such a proposal is meaningless without interesting examples,
string orbifolds appear to offer the first nontrivial examples.

It should also be said that 
there exists a group of mathematicians who already use
quotient stacks to describe string orbifolds.
Indeed, as described above, quotient stacks are an overcomplicated way
to describe group actions on covering spaces, and also possess
extra structure that gives them an interpreation as a sort of
`generalized space,' so that, for example, one can make sense of
differential geometry on a quotient stack.  However, as far as the
author has been able to determine, the mathematicians in question
have not done any of the work required to justify the claim
that a string orbifold CFT coincides with the CFT for a string 
compactified
on a quotient stack, or even to justify the claim that the notion of
string compactification on a stack makes sense.  They do not seem to
have attempted to study sigma models on stacks, and do not even realize
why this is relevant.  They also do not seem to be aware of even the most
elementary physical implications of such a claim, such as the fact that,
to be consistent, any deformations of a string orbifold CFT would have
to be interpreted in terms of deformations of the quotient stack,
rather than the quotient space.  Thus, mathematicians reading this paper
should interpret our work as the beginnings of a program to fill in
the logical steps that they seem to
have omitted.  In fact, this paper was written
for a physics audience; mathematicians are encouraged to instead read
lecture notes \cite{wisc} we shall publish shortly.

\section{String orbifolds do not live on $X/\Gamma$}

During the preparation of \cite{qstx}, we had numerous conversations
with physicists claiming that string orbifolds describe
strings on quotient spaces.  As a result, we feel compelled
to spend some time examining this more closely.
Of course, no one would claim that a string orbifold is the
same thing as a sigma model on a quotient space, if for
no other reason than the fact that one sums over maps into the
cover, instead of the quotient.  However, some might argue
that this is merely an artifact of the description, and to
counter such arguments, we shall spend a bit of time examining
the issue more closely.  In the process, we shall also gain some perspective
that will be useful later.

Given any space $X$, we can define a category of
maps into $X$.  Specifically,
\begin{enumerate}
\item Objects in this category are continuous maps $f: Y \rightarrow X$
from any other topological space $Y$ into $X$.
\item Morphisms $( Y_1 \stackrel{f_1}{\longrightarrow} X ) \:
\longrightarrow \: ( Y_2 \stackrel{f_2}{\longrightarrow} X )$
are continuous maps $\lambda: Y_1 \rightarrow Y_2$ such that the
following diagram commutes:
\begin{displaymath}
\begin{array}{ccc}
Y_1 & \stackrel{\lambda}{\longrightarrow} & Y_2 \\
\makebox[0pt][r]{ $\scriptstyle{ f_1 }$ } \downarrow & &
\downarrow \makebox[0pt][l]{ $\scriptstyle{ f_2 }$ } \\
X & = & X
\end{array}
\end{displaymath}
\end{enumerate}

It is a standard result that a topological space $X$ is determined
by this category.  In other words, if we
know all of the maps into a space, we can reconstruct the space.

Let us consider string orbifolds.  Can string orbifolds be
literally interpreted as sigma models on quotient spaces?
Not unless the orbifold group $\Gamma$ acts freely.

The most efficient way to study this problem is to first describe
the twisted sector maps more elegantly.
Instead of talking about maps from twisted sectors into $X$,
as in figure~\ref{figsig1},
an equivalent and more elegant description is as a pair
\begin{displaymath}
\left( \, E \: \stackrel{\pi}{\longrightarrow} \: \Sigma, \:
E \: \stackrel{f}{\longrightarrow} \: X \, \right)
\end{displaymath}
where\begin{tabbing}
abcdefg \= yuck \kill  
 \> $\Sigma$ is the string worldsheet, \\
 \> $E \rightarrow \Sigma$ is
a principal $\Gamma$-bundle, and \\
 \> $f: E \rightarrow X$ is a continuous $\Gamma$-equivariant map.
\end{tabbing}
In this description, a twisted sector is the same thing as an
equivalence class of bundles.  To see how to recover the original
description, simply restrict $E$ to a maximally-large contractible
open subset of $\Sigma$.  (If we describe the Riemann surface
$\Sigma$ in terms of a polygon with sides identified, such an open
set would be the interior of the polygon.)  Over a contractible open set,
$E$ is trivializable, so pick some section $s$.  The ordinary
description of maps from twisted sectors into $X$ is precisely
the action of $f$ on that section $s$, and the group elements acting
at the boundary determine to what extent $s$ fails to be a global section.
Precisely because $f: E \rightarrow X$ is $\Gamma$-equivariant,
knowing the action of $f$ on such a section $s$ completely determines $f$,
i.e., there is no extra information contained in the map $f: E \rightarrow
X$ that is not also contained in the twisted sector map into $X$. 

Given such a pair $( E \stackrel{\pi}{\longrightarrow} \Sigma,
E \stackrel{f}{\longrightarrow} X )$, we can derive a continuous
map $g: \Sigma \rightarrow X/\Gamma$ into the quotient space.
Specifically, for any point $y \in \Sigma$, let $e$ be any point
in the fiber of $E$ over $y$.  If we denote the canonical projection
$X \rightarrow X/\Gamma$ by $\pi_0$, then define $g: \Sigma \rightarrow
X/\Gamma$ by, $g(y) = (\pi_0 \circ f)(e)$.  It is straightforward to
check that this is well-defined and continuous.

To what extent is a pair $(E \stackrel{\pi}{\longrightarrow}
\Sigma, E \stackrel{f}{\longrightarrow} X)$ the same thing as a map
$\Sigma \rightarrow X/\Gamma$?  Given the former, we can construct
the latter; however, in a well-defined sense, if $\Gamma$
does not act freely, then there are more maps of the former form
than of the latter.  In other words, 
pairs $( E \stackrel{\pi}{\longrightarrow} \Sigma, 
E \stackrel{f}{\longrightarrow} X)$ seem to be defining maps
into a space related to $X/\Gamma$ but containing ``extra
information'' over any singularities.

In the next section, we shall argue that a string orbifold,
although not quite a sigma model on a quotient space $X/\Gamma$,
appears to be
literally a sigma model on something called a quotient {\it stack}
$[X/\Gamma]$.  (We shall argue this first by noting that
a sum over maps into a ``quotient stack'' duplicates both 
the sum over twisted sectors and the functional integral within
each twisted sector, a smoking gun for a sigma model interpretation,
and then propose a classical action for a sigma model on a stack.)

What is a quotient stack?
To be very brief, a quotient stack is an example of what is sometimes
known as a ``generalized space.'' For example, instead of possessing
a set of points, it has a {\it category} of points.
In general, such spaces can be defined in terms of the category
of maps into them.  This is somewhat analogous to noncommutative
geometry, where spaces are defined by the algebra of functions on them.
Here, instead of working with the algebra of functions on the space, one works
with the category of continuous maps into the space.
This may sound somewhat cumbersome, but it is actually an ideal setup
for string sigma models.

In passing, we should also mention that the idea of defining
spaces in terms of the maps into them is a commonly-used setup
in algebraic geometry (see discussions of ``Grothendieck's functor
of points'' in, for example, \cite[section II.6]{mumred} or
\cite[section VI]{eisenharris}).

How are quotient stacks related to quotient spaces?
For example, when the orbifold group $\Gamma$ acts freely,
the quotient stack $[X/\Gamma]$ and the quotient space
$X/\Gamma$ are homeomorphic.  So, one way of thinking about
quotient stacks is that they look like quotient spaces,
except that they have some ``extra structure'' over the singularities.
That ``extra structure'' makes quotient stacks much better behaved
than quotient spaces, and is responsible for features of string
orbifolds than people have labelled ``stringy'' in the past.

\section{Unravel definitions -- path integral sums}

Quotient stacks $[X/\Gamma]$ are defined by the category of maps
from all topological spaces into $[X/\Gamma]$.
In particular, a continuous map from any topological space $\Sigma$ 
into $[X/\Gamma]$ is a pair 
\begin{displaymath}
\left( \, E \: \stackrel{\pi}{\longrightarrow} \: \Sigma, \:
E \: \stackrel{f}{\longrightarrow} \: X \, \right)
\end{displaymath}
where
\begin{tabbing}
abcdefg \= yuck \kill  
 \> $E \rightarrow \Sigma$ is
a principal $\Gamma$-bundle, and \\
 \> $f: E \rightarrow X$ is a continuous $\Gamma$-equivariant map.
\end{tabbing}

But, we argued earlier that these are precisely the things one
literally sums over in a string orbifold -- by summing over maps
into a quotient stack, one duplicates both the twisted sector sum
as well as the sum over maps within each twisted sector.
So, in other words, a string orbifold is {\it literally} a sum
over maps from the worldsheet into the quotient stack $[X/\Gamma]$,
a `smoking gun' for an interpretation as a sigma model.
Put another way, after unraveling definitions, string orbifolds
appear to be sigma models on quotient stacks,
at least judging by the path integral sum.  In the next section,
we shall describe a proposal for a classical action for a sigma model
on a stack, and further justify this conclusion.

Some readers might find this conclusion to be somewhat extreme -- 
after all, physicists
have not explicitly considered stacks in the past.  
Indeed, part of the point of \cite{qstx} was to set up the basics
required for physicists to begin to consider string compactification
on stacks.  One needs some nontrivial examples of such compactifications
to begin to take such a notion seriously, and string orbifolds
appear to offer such an example.


\section{Classical actions for sigma models on stacks}

Now, just because a string orbifold is literally a sum over
maps into $[X/\Gamma]$ does not itself imply that a string
orbifold is necessarily a sigma model on $[X/\Gamma]$ -- for that
to be the case, we also need the action associated to each map
to be the same, so that the path integral is the same {\it weighted}
sum over maps.  
In this section we shall describe a natural proposal for a classical
action for a sigma model on a stack, and check that this not only
duplicates standard sigma models when the target stack is an honest space,
but also duplicates string orbifolds (down to the $|G|^{-1}$ factors
in partition functions, when the target is a global quotient stack.


Let ${\cal F}$ be a stack, with atlas $X$.
(We shall only attempt to describe sigma models on stacks with
atlases.)  For readers not well-acquainted with stacks,
for $X$ to be an atlas for ${\cal F}$ implies that
\begin{itemize}
\item implicitly there is also a fixed map $X \rightarrow {\cal F}$
(which is required to be a surjective local homeomorphism)
\item for any space $Y$ and map $Y \rightarrow {\cal F}$,
the fibered product $Y \times_{ {\cal F} } X$ is an honest space, not a stack.
\end{itemize}

For example, if ${\cal F}$ is a space, not just a stack
(spaces are special cases of stacks), then ${\cal F}$ is its own atlas,
and $Y \times_{ {\cal F} } X = Y$ for any $Y$.
For another example, suppose ${\cal F}$ is a quotient stack $[X/G]$,
with $G$ discrete and acting by diffeomorphisms on a smooth space $X$.
In such a case, $X$ is an atlas for $[X/G]$.  In this case,
$Y \times_{ [X/G] } X$ is a principal $G$-bundle over $Y$,
partially specifying the map $Y \rightarrow [X/G]$,
and the projection map $Y \times_{ [X/G] } X \longrightarrow X$
is the $G$-equivariant map from the total space of the bundle to $X$,
specifying the rest of the map $Y \rightarrow [X/G]$.

Now, the natural description of
a sigma model with target ${\cal F}$, formulated on
(base) space $Y$, is a sum over equivalence classes\footnote{A sigma model
path integral is a sum over maps, after all, hence one must take
equivalence classes in order to make sense out of such a sum.} of maps
$Y \rightarrow {\cal F}$, weighted by
$\exp(iS)$, where the classical action $S$ is
formulated as follows.  Fix a map $\phi: Y \rightarrow
{\cal F}$.  If we let\footnote{
Note that since both $Y \times_{ {\cal F} } X$ and $X$ are ordinary
spaces, $\Phi$ is a map in the ordinary sense of the term.}
$\Phi: Y \times_{ {\cal F} } X \rightarrow X$ denote the
second projection map (implicitly encoding part of the map
$\phi: Y \rightarrow {\cal F}$), then the natural proposal for
the bosonic part of the classical
action for a sigma model on ${\cal F}$ is given by \cite{qstx}
\begin{equation}
\int d^2 \sigma \, \left( \pi_1^* \phi^* G_{\mu \nu} \right)
\pi_1^* h^{\alpha \beta} 
\left( \frac{ \partial \Phi^{\mu} }{ \partial \sigma^{\alpha} } \right)
\left( \frac{ \partial \Phi^{\nu} }{ \partial \sigma^{\beta} } \right)
\end{equation}
where $h^{\alpha \beta}$ is the worldsheet metric, $\phi^* G$ denotes
the pullback of the metric on ${\cal F}$ to $Y$ (metrics on ${\cal F}$
are described
in terms of their pullbacks), $\pi_1: Y \times_{ {\cal F} } X \rightarrow
Y$ is the projection map, and this action is integrated over
a lift\footnote{Sensible essentially because
the (projection) map $\pi_1: Y \times_{ {\cal F} } X \rightarrow Y$
is a surjective local homeomorphism.}
of $Y$ to $Y \times_{ {\cal F} } X$.

A few examples should help clarify this description:
\begin{enumerate}
\item Suppose ${\cal F}$ is an ordinary space.
Then the path integral is a sum over maps into that space,
and as $Y \times_{ {\cal F} } X = Y$ (taking the atlas $X$ to be
${\cal F}$ itself), we see that $\Phi = \phi$,
and so 
in this case the classical action proposed
above duplicates the usual classical action, 
\begin{displaymath}   
S \: \sim \: \int d^2 \sigma \left( \phi^* G_{\mu \nu} \right)
h^{\alpha \beta}
\frac{ \partial \phi^{\mu} }{ \partial \sigma^{\alpha} } 
\frac{ \partial \phi^{\nu} }{ \partial \sigma^{\beta} } 
\: + \: \cdots
\end{displaymath}
as well as the path integral sum.
Thus, the description above duplicates sigma models
on ordinary spaces.
\item Suppose ${\cal F} = [X/G]$, where $X$ is smooth and
$G$ is a nontrivial action of a discrete group by diffeomorphisms.
Then the path integral is a sum over equivalence classes of maps
$Y \rightarrow [X/G]$, which is to say, equivalence classes of
principal $G$-bundles on $Y$ together with $G$-equivariant maps
from the total space of the bundle into $X$.  It is easy to check
that the proposed classical action above duplicates the usual
classical action for a string orbifold.  Also, by summing over
(equivalence classes of) maps $Y \rightarrow {\cal F}$,
note we are summing over both twisted sectors as well
as maps within any given twisted sector.

Now, for each such map $\phi: Y \rightarrow {\cal F}$, 
there are $| G |$ lifts of $Y$ to the 
total space of the bundle (which is $Y \times_{ [X/G] } X$),
{\it i.e.}, $| G |$ twisted sector maps, as they usually appear in
physics.

Note we are only summing over equivalence classes of bundles,
not all possible twisted sector maps.  However, we can trivially
sum over all possible twisted sector maps, at the cost of
overcounting by $| G |$.  Hence, we can equivalently describe
this in terms of a sum over twisted sector maps, 
but weighted by $| G |^{-1}$.  Hence we recover both the path
integral sum and the overall multiplicative factor of $|G|^{-1}$
appearing in string orbifold partition functions, for example,
the one-loop partition function
\begin{displaymath}    
Z(T^2) \: = \: \frac{1}{|G|}
\sum_{\begin{array}{c} \scriptstyle{ g,h \in G } \\
\scriptstyle{ gh=hg}  \end{array}
 } Z_{(g,h)}
\end{displaymath}
\end{enumerate}

Thus, we see that the natural definition of a sigma model on a stack
duplicates not only sigma models on ordinary spaces,
but also string orbifolds when the target is a quotient stack, 
even down to the $| G |^{-1}$ factor
appearing in partition functions.

We should take this opportunity to also note that this description
of sigma models does not make any assumptions concerning the dimension
of the base space $Y$ -- classically there are analogues of `string' orbifolds
in every dimension, all obtained precisely by gauging the action of
a discrete group on the target space of a sigma model.

So far we have only recovered known results; let us now try 
something new.  Suppose the target ${\cal F}$ is a gerbe.
For simplicity, we shall assume that ${\cal F}$ is the
canonical trivial $G$-gerbe on a space $X$.
Such a gerbe is described by the quotient stack $[X/G]$,
where the action of $G$ on $X$ is trivial.
Using the notion of sigma model on a stack as above,
one quickly finds that the path integral for this target space
is the same as the path integral for a sigma model on $X$,
up to an overall multiplicative factor (equal to the number of
equivalence classes of principal $G$-bundles on $Y$).
As overall factors are irrelevant in path integrals,
the result appears to be that a string on the canonical trivial
gerbe is the same as a string on the underlying space.
More generally, it is natural to conjecture that
strings on flat gerbes are equivalent to
strings on underlying spaces, but with flat $B$ fields.
In particular, such a result would nicely dovetail with the well-known
fact that
a coherent sheaf on a flat gerbe is equivalent to a `twisted' sheaf
on the underlying space, the same twisting that occurs in the presence
of a $B$ field.  (For physicists, this is an alternative to the
description in terms of modules over Azumaya algebras that has
recently been popularized \cite{kapustin}.)

So far we have only discussed classical actions for sigma models
on stacks, but there is much more that must be done before 
one can verify that the notion of a sigma model on a stack
is necessarily sensible.  In effect, we have only considered
local behavior, but in order to be sure this notion is sensible
after quantization, one also needs to consider global phenomena.
Such considerations were the source of much hand-wringing when
nonlinear sigma models on ordinary spaces were first introduced
(see for example \cite{mmn}), and must be repeated for stacks.

\section{More results}

So far we have argued that a string orbifold is literally
a sigma model on a quotient {\it stack}, as opposed to a quotient space.
What does this do for us?  We just saw that this naturally explained
the structure of twisted sectors; next we shall outline how this
also gives geometric perspectives on many other features of
string orbifolds, previously considered ``stringy.''

Specifically:
\begin{enumerate}
\item Smoothness.  String orbifold CFT's do not suffer from any singularities;
they behave as if they were sigma models on smooth spaces.
This led to the old lore that ``strings smooth out singularities.''
However, quotient stacks $[X/\Gamma]$ are always smooth
(for $X$ smooth and $\Gamma$ acting by diffeomorphisms),
and so one naturally expects that a sigma model on $[X/\Gamma]$ should
always be well-behaved.  The old lore ``strings smooth out singularities''
is merely a consequence of misunderstanding the target space geometry;
nothing ``stringy'' is really involved.
\item B fields.  Another often-quoted fact concerning string
orbifolds is that the B field should somehow have nonzero holonomy
about shrunken exceptional divisors 
\cite{paulz2,douglaszn,sarah,katrin1,edstrings95}.
Understanding string orbifolds as sigma models on quotient stacks
gives a natural understanding of what is meant by such claims.
Specifically, the ``extra information'' contained by a quotient stack
over singularities of the quotient space is a gerbe\footnote{A gerbe
is a formal structure corresponding to B fields, just as bundles
correspond to gauge fields.}, that precisely
duplicates standard results on B fields at quotient singularities.
A gerbe is, after all, just a special kind of stack, so the reader
should not be surprised to find information about B fields
given in the geometry of stacks. 
\item The role of equivariance.  Bundles and sheaves on $[X/\Gamma]$
are the same thing as $\Gamma$-equivariant bundles and sheaves on $X$.
So, we now gain a new perspective on the role that equivariance plays in
describing fields on string orbifolds.
\item Twist fields.  Ordinarily the low-energy spectrum of a 
string compactification is determined by the cohomology of the target
space.  However, we argue in \cite{qstx} that for stacks
matters are slightly more interesting, and the low-energy spectrum
is determined by the cohomology of an auxiliary stack,
known as the associated inertia group stack $I_{[X/\Gamma]}$.
(This result appears very obscure to most physicists, many of whom
have traditionally expected an understanding of twist fields
in terms of a cohomology of the quotient {\it space};
mathematicians familiar with stacks, on the other hand, will recognize
this result.)
This stack differs from $[X/\Gamma]$ precisely when the quotient
stack cannot be understood as an ordinary space.
More to the point, $I_{[X/\Gamma]}$ naturally encodes the ``twists''
that give twist fields their name, and in fact has the form
\cite{toen}
\begin{displaymath}
I_{[X/\Gamma]} \: \cong \: \coprod_{[g]} \, \left[ \, X^g / C(g) \, \right]
\end{displaymath}
which of course compares well with orbifold Euler characteristics
\cite{hirzhofer}
\begin{displaymath}
\chi_{orb}(X, \Gamma) \: = \: \sum_{[g]} \, \chi( X^g / C(g) )
\end{displaymath}
where $[g]$ denotes conjugacy classes of $\Gamma$ and
$C(g)$ the centralizer of $g \in \Gamma$.
\end{enumerate}

In \cite{qstx} we work through the points above in much greater detail,
and also describe how some other features of string orbifolds are
clarified.

So far, we have argued that understanding string orbifolds
as sigma models on quotient stacks gives a natural geometric
explanation to
many features of string orbifolds that were previously considered
``inherently stringy.'' 

But what new things can we do with quotient stacks?
We shall list a few general directions:
\begin{enumerate}
\item Deformation theory.  In constructing moduli spaces of string
vacua, physicists have assumed string orbifolds describe strings
on quotient spaces.  If string orbifolds actually describe
strings on quotient stacks, then these old arguments must be
reconsidered -- for example, a blowup modulus must be interpreted
as some hypothetical K\"ahler modulus of the quotient stack, and not 
in terms of a (partial) resolution of the quotient space.
We have given indirect evidence in \cite{qstx} that such considerations
might ultimately be equivalent to working with a resolution with
a nonzero B field on the exceptional divisor, but a tremendous amount
of work remains to be done to check whether this is actually the case.
\item New string compactifications.  There are more stacks
than just global quotient stacks.  So, given our preliminary work
on string compactification on stacks (bolstered by the nontrivial
example of string orbifolds), 
one could begin seriously
studying compactifications on other stacks.
\item M-theory\footnote{We use ``M-theory'' in the sense of,
quantum theory underlying eleven-dimensional supergravity,
as opposed to some hypothetical master theory.}
orbifolds.  If one were careful, in the past one could have
objected that orbifolds in M-theory could hardly be considered
well-understood.  After all, orbifolds in string theory possess
twisted sectors, twist fields, and many other features which
naively seemed to be ``inherently stringy'' -- it was not at all
clear how one could make sense of such things in M-theory.
Now, however, we can understand these matters better.
We can define an M-theory orbifold to be, M-theory compactified on
a quotient stack.  Then, for example, membranes on a quotient
stack have an obvious twisted-sector-type structure.
One might be able to do quite a bit more with this description -- for example,
it may be possible to directly understand the Horava-Witten
\cite{petred} $E_8$ multiplets as arising naturally, whereas
in their original description these multiplets had to be added in
manually.
\end{enumerate}

\section{Conclusions}

In this short note we have outlined some recent results on
string orbifolds.  Specifically, we have outlined how
string orbifolds seem to be literally sigma models on quotient stacks
(not quotient spaces).
We have also described how, from this new perspective, many properties of
string orbifolds that were formerly considered ``stringy,''
actually seem to have a simple geometric understanding in terms
of the target
space geometry.  

However, a tremendous amount of work remains to be done,
not only to better understand what it means to compactify a string
on a stack (indeed, to check whether this is indeed a sensible notion),
but also to understand the consequences of this perspective
(such as deformation theory of string orbifolds).
In the conclusions of \cite{qstx} we
have given a lengthy list of further topics that could be pursued.

\section{Acknowledgements}

We would like to thank P.~Aspinwall, D.~Ben-Zvi, S.~Dean, J.~Distler, T.~Gomez,
R.~Hain,
S.~Katz, A.~Knutson, D.~Morrison,
B.~Pardon, R.~Plesser, and M.~Stern for useful conversations.
In particular, we would especially like to thank T.~Gomez for extensive
communications concerning quotient stacks in particular, A.~Knutson for
numerous communications regarding related general mathematics, and
D.~Ben-Zvi for originally pointing out that the twisted ``bundles''
on D-brane worldvolumes should be understandable in terms of bundles
on stacks, the observation which ultimately led to this work.

\end{document}